\documentclass[twocolumn,prl,twocolumn,superscriptaddress]{revtex4-1}
\usepackage[latin9]{inputenc}
\usepackage{color}
\usepackage{amsmath}
\usepackage{amssymb}
\usepackage{graphicx}
\usepackage{tikz}
\usepackage{amsmath,bm,amssymb,latexsym,galois,amsthm,euscript,dsfont}
\usepackage[all,cmtip]{xy}
\usepackage{dsfont} 
\usepackage{mathrsfs}

\usepackage[unicode=true,
 bookmarks=true,bookmarksnumbered=false,bookmarksopen=false,
 breaklinks=false,pdfborder={0 0 1},backref=false,colorlinks=true]{hyperref}
\hypersetup{linkcolor=magenta, urlcolor=blue, citecolor=BLUE, pdfstartview={FitH}, hyperfootnotes=false, unicode=true}

\makeatletter
\@ifundefined{textcolor}{}
{%
 \definecolor{BLACK}{gray}{0}
 \definecolor{WHITE}{gray}{1}
 \definecolor{RED}{rgb}{1,0,0}
 \definecolor{GREEN}{rgb}{0,1,0}
 \definecolor{BLUE}{rgb}{0,0,1}
 \definecolor{CYAN}{cmyk}{1,0,0,0}
 \definecolor{MAGENTA}{cmyk}{0,1,0,0}
 \definecolor{YELLOW}{cmyk}{0,0,1,0}
}

\usepackage{amsfonts}\usepackage{tabularx}\usepackage{dcolumn}\usepackage{bm}\usepackage{graphicx}\usepackage{epstopdf}

\hypersetup{
    colorlinks,
    linkcolor={red!50!black},
    citecolor={blue!90!black},
    urlcolor={blue!90!black}
}

\theoremstyle{definition}
\newtheorem{defn}{Definition}

\newtheorem{prop}[defn]{Proposition}
\newtheorem{lemma}[defn]{Lemma}

\newtheorem*{defn*}{Definition}
\newtheorem{thm}{Theorem}
\theoremstyle{remark}
\newtheorem*{pf}{PROOF}

\newcommand\EC{\EuScript{C}}
\newcommand\EP{\EuScript{P}}
\newcommand\ES{\EuScript{S}}

\makeatother

\begin{document}

\widetext


\title{Hierarchy of Genuine Multipartite Quantum Correlations}
\author{Zhih-Ahn Jia}
\email{giannjia@foxmail.com}
\affiliation{Key Laboratory of Quantum Information, Chinese Academy of Sciences, School of Physics, University of Science and Technology of China, Hefei, Anhui, 230026, P. R. China}
\affiliation{Synergetic Innovation Center of Quantum Information and Quantum Physics, University of Science and Technology of China, Hefei, Anhui, 230026, P. R. China}
\author{Rui Zhai}
\affiliation{Institute of Technical Physics, Department of Engineering Physics, Tsinghua University, Beijing 10084, China}

\author{Shang Yu}
\affiliation{Key Laboratory of Quantum Information, Chinese Academy of Sciences, School of Physics, University of Science and Technology of China, Hefei, Anhui, 230026, P. R. China}
\affiliation{Synergetic Innovation Center of Quantum Information and Quantum Physics, University of Science and Technology of China, Hefei, Anhui, 230026, P. R. China}

\author{Yu-Chun Wu}
\email{wuyuchun@ustc.edu.cn}
\affiliation{Key Laboratory of Quantum Information, Chinese Academy of Sciences, School of Physics, University of Science and Technology of China, Hefei, Anhui, 230026, P. R. China}
\affiliation{Synergetic Innovation Center of Quantum Information and Quantum Physics, University of Science and Technology of China, Hefei, Anhui, 230026, P. R. China}

\author{Guang-Can Guo}\affiliation{Key Laboratory of Quantum Information, Chinese Academy of Sciences, School of Physics, University of Science and Technology of China, Hefei, Anhui, 230026, P. R. China}
\affiliation{Synergetic Innovation Center of Quantum Information and Quantum Physics, University of Science and Technology of China, Hefei, Anhui, 230026, P. R. China}

\date{\today}

\begin{abstract}
Classifying states which exhibiting different statistical correlations is among the most important problems in quantum information science and quantum many-body physics. In bipartite case, there is a clear hierarchy of states with different correlations: total correlation (T) $\supsetneq$ discord (D) $\supsetneq$ entanglement (E) $\supsetneq$ steering (S) $\supsetneq$ Bell~nonlocality (NL). However, very little is known about genuine multipartite correlations (GM$\mathcal{C}$) for both conceptual and technical difficulties. In this work, we show that, for any $N$-partite qudit states, there also exist such a hierarchy: genuine multipartite total correlations (GMT) $\supseteq$ genuine multipartite discord (GMD) $\supseteq$ genuine multipartite entanglement (GME) $\supseteq$ genuine multipartite steering (GMS) $\supseteq$ genuine multipartite nonlocality (GMNL). Furthermore, by constructing precise states, we show that GMT, GME and GMS are inequivalent with each other, thus GMT $\supsetneq$ GME $\supsetneq$ GMS.
\end{abstract}


\maketitle


\emph{Introduction.}\textemdash
Investigating the nature of different kinds of correlations is one of the central problems in quantum information theory \cite{preskill1998,Nielsen2010} and quantum many-body physics \cite{Wen2004,Zeng2015}, besides its theoretical interest, it also has a crucial applicative importance. Entanglement \cite{Amico2008,Horodecki2009,Guhne2009} and Bell nonlocality \cite{Brunner2014bell} are two prototypical examples, which are among the most striking features of quantum mechanics and they are also important resources for information tasks. The landmark works of Einstein \emph{et al.} \cite{einstein1935} and Bell \cite{bell1987einsteinpodolskyrosen} indicate that quantum theory is beyond the local hidden variable (LHV) model , the phenomenon has been designated as nonlocality. Since then, many new kind of correlations have also been extensively investigated, for example, total correlation and quantum discord \cite{Henderson2001,Ollivier2001,Zurek2003,Modi2010,Modi2012} which is the measure of nonclassical correlations between two subsystems of a physical system. It includes correlations that are due to quantum physical effects but do not necessarily involve quantum entanglement \cite{Henderson2001,Ollivier2001,Zurek2003,Modi2010,Modi2012}. The other one is quantum steering \cite{Reid2009} which is initially introduced by Schr\"{o}dinger \cite{schrodinger1935discussion} and recently rigourously defined by Wiseman \emph{et al.} \cite{Wiseman2007} to describe the ability of one experimenter Alice to remotely prepare an ensemble of states for another experimenter Bob by performing a local measurement on her half of bipartite system and communicating the results to Bob. The steerability is characterized by the failure of local hidden states (LHS) model. Unlike entanglement and nonlocality, quantum steering and discord are intrinsically asymmetric \cite{Zurek2003,Bowles2014}.

Having been confronted with the observation that every pure entangled state do violate some Bell inequality, during a long period of time, it is believed that entanglement and nonlocality  are the different aspects of the same physical phenomenon. It was Werner who first indicated the inequivalence between entanglement and nonlocality by giving the explicit examples of mixed entangled states (now named as Werner states) which do not exhibit nonlocal properties under projective measurements \cite{wener1989quantum}. There are many known examples of entangled states admitting LHS model for projective measurements \cite{wener1989quantum,Wiseman2007,Almeida2007,Bowles2014,Jevtic2014}, examples for positive operator-valued measurements (POVMs) was provided by Barrett \cite{Barrett2002}. Recently, Quintino \emph{et al.} \cite{Quintino2015} provided a general way to construct such kind of entangled but unsteerable (under POVMs) states from ones under projective measurements, they also constructed some examples of steerable but local (both under POVMs) states (see also Ref. \cite{Hirsch2013}). Thus we have a clear hierarchy of the bipartite qudit state set $\mathcal{S}^{d,2}$ as $\mathcal{S}^{d,2}_{T}\supsetneq\mathcal{S}^{d,2}_{D}\supsetneq \mathcal{S}^{d,2}_{E}\supsetneq \mathcal{S}^{d,2}_{S}\supsetneq \mathcal{S}^{d,2}_{NL}$ \cite{wener1989quantum,Barrett2002,Zurek2003,Wiseman2007,Bowles2014,Quintino2015}, hereinafter we denote the set of $N$-partite $d$-dimensional quantum state (which exhibiting $\mathcal{C}$-correlation) as $\mathcal{S}^{d,N}$ ($\mathcal{S}_{\mathcal{C}}^{d,N}$), i.e., total correlation, quantum discord, entanglement, steering and nonlocality are inequivalent notions for bipartite system. But for multipartite case, things become much subtler and related problems still largely remain open \cite{Zurek2003,Amico2008,Horodecki2009,Modi2012,Brunner2014bell}.

In a work \cite{Svetlichny1987distinguishing} in 1987, Svetlichny raised the question whether quantum mechanics exhibits genuine multipartite nonlocality (GMNL) in the sense that this kind of nonlocality can not be simulated using any nonlocality that only exists in some subsystems. The answer is \emph{yes}, it has been experimentally verified that our nature do exhibit genuine nonlocality \cite{lavoie2009,Lu2011}. Actually, our nature do exhibit genuine multipartite $\mathcal{C}$ (GM$\mathcal{C}$) correlation, where $\mathcal{C}$=T, D, E, S, NL, since Greenberger-Horne-Zeilinger state \cite{greenberger1989going} is a typical example of such kind of states. It is trivial to construct multipartite state which can indicate the inequivalence of, for example entanglement and nonlocality, since we can take product of a two partite state which indicates this kind of inequivalence and so arbitrary state. Nevertheless, we can naturally ask if, for any number of parties $N$, we still have a precise hierarchy of $N$-partite state set which exhibit GM$\mathcal{C}$-correlation. Actually it is a quite challenging problem. Augusiak \emph{et al.} indicates the inequivalence of GME and GMNL \cite{Augusiak2015}, there are also some works concentrated on constructing LHV model for entanglement states \cite{Cavalcanti2016,Hirsch2016}, but very little is know about the relations of other $\mathrm{GM\mathcal{C}}$.

\begin{figure}
\includegraphics[scale=1.0]{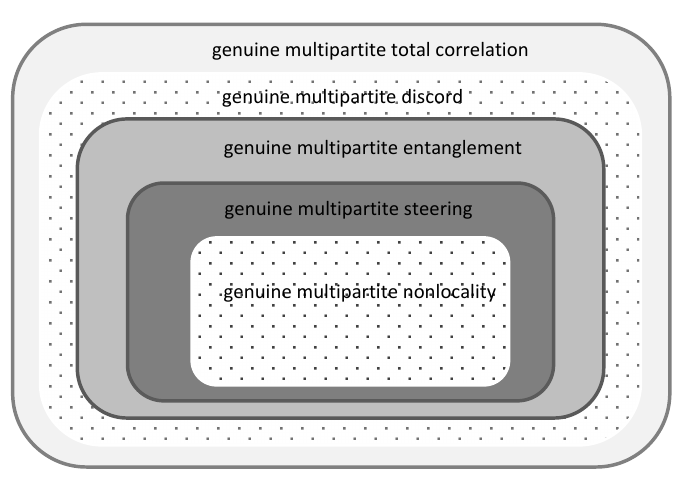}
\caption{\label{fig:correlation} (color online). The hierarchy of genuine multipartite correlations $\mathcal{S}^{d,N}_{GMT}\supseteq \mathcal{S}^{d,N}_{GMD}\supseteq \mathcal{S}^{d,N}_{GME}\supseteq \mathcal{S}^{d,N}_{GMS}\supseteq \mathcal{S}^{d,N}_{GMNL}$ and $\mathcal{S}^{d,N}_{GMT}\supsetneq \mathcal{S}^{d,N}_{GME}\supsetneq \mathcal{S}^{d,N}_{GMS}$.}
\end{figure}

We denote the set of $d$-dimensional $N$-partite states which exhibiting $\mathcal{C}$ (resp. GM$\mathcal{C}$) correlations as $\mathcal{S}_{\mathcal{C}}^{d,N}$ (resp. $\mathcal{S}_{GM\mathcal{C}}^{d,N}$). For pure bipartite state set $\mathcal{S}^{d,2}_{pure}$, we have $\mathcal{S}^{d,2}_{pure T}=\mathcal{S}^{d,2}_{pure D}=\mathcal{S}^{d,2}_{pure E}= \mathcal{S}^{d,2}_{pure  S}= \mathcal{S}^{d,2}_{pure NL}$. But for general mixed states, there is a clear hierarchy, $\mathcal{S}^{d,2}_{T}\supsetneq\mathcal{S}^{d,2}_{D}\supsetneq\mathcal{S}^{d,2}_{E}\supsetneq \mathcal{S}^{d,2}_{S}\supsetneq \mathcal{S}^{d,2}_{NL}$, which can been shown by explicit construction of states intermediate between two state sets \cite{wener1989quantum,Barrett2002,Wiseman2007,Bowles2014,Quintino2015}. With the same spirit, we investigate the GM$\mathcal{C}$. The main result we will show in this paper is the following theorem:

\begin{thm}\label{thm:thm1} For any number of parties $N$, we have an inclusion relation of qudit state sets $\mathcal{S}^{d,N}_{GMT}\supseteq \mathcal{S}^{d,N}_{GMD}\supseteq \mathcal{S}^{d,N}_{GME}\supseteq \mathcal{S}^{d,N}_{GMS}\supseteq \mathcal{S}^{d,N}_{GMNL}$ and $\mathcal{S}^{d,N}_{GMT}\supsetneq \mathcal{S}^{d,N}_{GME}\supsetneq \mathcal{S}^{d,N}_{GMS}$.
\end{thm}
The result has been depicted in the Fig. \ref{fig:correlation}, this theorem gives an almost complete hierarchy of GM$\mathcal{C}$.

\emph{Inequivalences of GME, GMS and GMNL.}\textemdash
Consider $N$ parties $S=\{1,\cdots,N\}$ sharing an $N$-partite qudit state $\rho\in \mathcal{S}^{d,N}$, an $n$-partition of $S$ is a partition of $S$ with $n$ disjoint components $S_1|S_2|\cdots|S_n$ such that $S=\sqcup_{j=1}^n S_j$, we denote the set of all $n$-partition with $\mathscr{P}_n$, e.g., for three-party system $S=\{1,2,3\}$,  $\mathscr{P}_2=\{1|23,12|3,13|2\}$. A 2-partition is generally named as bipartition and is often denoted with $A|A^c$ where $A^c$ denotes the complement of $A$, there are $(2^{N-1}-1)$ kinds of bipartition for an $N$-partite system. We the denote the set of all partitions as $\mathscr{P}=\sqcup_m \mathscr{P}_m$.

The main point of the definition of GMNL state is that it falsify the hybrid LHV model \cite{Svetlichny1987distinguishing,Collins2002bell,Seevinck2002bell,Bancal2011detecting}: for $N$ parties $\{1,\cdots,N\}$, and any set of measurements $x_i\in\mathfrak{M}_1,\cdots,x_N\in\mathfrak{M}_N$, the statistics satisfy
\begin{equation}\label{nl}
\begin{split}
&p(a_1,\cdots,a_N|x_1,\cdots,x_N)_{\rho} \\
=&\sum_{P\in \mathscr{P}_2}\wp_P \int d\lambda p(\lambda) p(a|x_A,\lambda)p(b|x_{A^c},\lambda),\nonumber
\end{split}
\end{equation}
where $\sum_{P\in \mathscr{P}_2}\wp_P=1$, and $\lambda$ is hidden variable. Note that here we only demand the probability factorize under hidden variable into two part, but this is already enough to define a GMNL state.

We say a state $\rho$ is hybrid separable if it can be written in forms of
\begin{align}\label{}
\rho=\sum_{P\in \mathscr{P}}\wp_{P}\rho_{P},\nonumber
\end{align}
where $\sum_{P\in \mathscr{P}}\wp_P=1$ and $\rho_{P}$ is a product state with respect to partition $P$, e.g, for tripartite partition $1|23$, it is $\sigma_i\otimes \sigma_{23}$. $\rho$ is a GME state if it falsifies the hybrid separable model.

In \cite{HeQY2013}, the notion of GMS is introduced. For clarity, we using three-partite genuine steering as an example. For three particle $\{1,2,3\}$ we have three bipartitions $\mathscr{P}_2=\{1|23,12|3,2|31\}$, there are six corresponding directed bipartitions $\mathscr{P}^*_2=\{1\lceil 23,1\rceil 23,\cdots\}$, thus the genuine steering is just the failure of the hybrid LHS model
\begin{align}\label{}
& p(a_1,\cdots,a_N)  \nonumber \\
& =\sum_{A\lceil A^c\in \mathscr{P}^*_2}\wp_{A\lceil A^c} \int d\lambda p(\lambda) p(a|x_A,\lambda)p_Q(b|x_{A^c},\lambda),\nonumber
\end{align}
where $\sum_{A\lceil A^c\in \mathscr{P}^*_2}\wp_{A\lceil A^c}=1$.

Actually there are still some very subtle things about the definition need to be made clear. For example, consider a genuine $N$-partite GMNL state $\rho$, the above definition require the failure of mixture of bidecomposition of measurement statistics, but for a given partition $P=A|A^{c}\in\mathscr{P}_2$, we never consider the correlation among the particles inside the  groups $A$ and $A^c$, we only concern about the correlation between $A$ and $A^c$, so there may exist correlations among the particles inside $A$ (or $A^c$) even beyond quantum mechanics. For example, it has been show that  with and without of this kind of restriction arrive at different definitions of GMNL \cite{Gallego2012,Bancal2013}. But here, for simplicity we will not concern such kind of differences.

For bipartite qudit state set $\mathcal{S}^{d,N}$, it has been show that GMNL and GME are inequivalent for any number of parties $N$ \cite{Augusiak2015}, since GMS falls in between GME and GMNL, thus we need to show that there exist some states which are GME but not GMS, and some other states which are GMS but not GMNL. Our construction of these two kinds of inequivalences is the same as the one of Augusiak \emph{et al.} \cite{Augusiak2015}, which is strongly dependent of the inequivalence of bipartite state sets. More precisely, starting from a given entangled bipartite state admitting LHS model under POVMs, we can construct an $N$-partite state which is unsteerable with respect to some bipartition but genuine entangled. Firstly, following the construction provided by Barrett \cite{Barrett2002}, we give the following proposition:
\begin{lemma}\label{prop:1}
Let $\mathcal{E}:B(\mathcal{H})\to B(\mathcal{H}')$ and $\mathcal{U}:B(\mathcal{K})\to B(\mathcal{K}')$ be two quantum channels, i.e., trace-preserving completely positive (TPCP) mappings, and $\rho_{AB}\in B(\mathcal{H}\otimes \mathcal{K})$ be a state, under the channel $\rho_{AB}\mapsto \tilde{\rho}_{A'B'}=\mathcal{E}\otimes \mathcal{U}(\rho_{AB})$ we have\\
  \indent (i) if $\rho_{AB}\in B(\mathcal{H}\otimes \mathcal{K})$ admits a LHV model for any POVMs $x_a\in \mathfrak{M}(\mathcal{H})$ and $y_b\in \mathfrak{M}(\mathcal{K})$ as  $p(a,b|x,y)=\int q(\lambda)p(a|x,\lambda)p(b|y,\lambda)d\lambda$, the state $\tilde{\rho}_{A'B'}$ must admit a LHV model with respect to the tensor product structure $\mathcal{H}'\otimes \mathcal{K}'$ ;\\
  \indent (ii) if $\rho_{AB}\in B(\mathcal{H}\otimes \mathcal{K})$ admits a LHS model for any POVMs $x_a\in \mathfrak{M}(\mathcal{H})$ and $y_b\in \mathfrak{M}(\mathcal{K})$ from $A$ to $B$ as $p(a,b|x,y)=\int q(\lambda)p(a|x,\lambda)\mathrm{tr}(y_b \rho_{\lambda})d\lambda$, the state $\tilde{\rho}_{A'B'}$ must admit a LHS model from $A'$ to $B'$ with respect to the tensor product structure $\mathcal{H}'\otimes \mathcal{K}'$;\\
  \indent (iii) if $\rho_{AB}\in B(\mathcal{H}\otimes \mathcal{K})$ admits a separable model $\rho_{AB}=\sum_{\lambda} q(\lambda)\sigma_{\lambda}\otimes\rho_{\lambda}$, the state $\tilde{\rho}_{A'B'}$ must admit a separable model with respect to the tensor product structure $\mathcal{H}'\otimes \mathcal{K}'$;
\begin{pf} We prove the most representative (ii), (i) and (iii) can be proved similarly (see also \cite{Augusiak2015} for proof of (i)). For any POVM measurements $x'_{a'}\in\mathfrak{M}(\mathcal{H}')$ and $y'_{b'}\in\mathfrak{M}(\mathcal{H}')$, the statistics
\begin{align}\label{}
p(a',b'|x',y') &= \mathrm{tr}[x'_{a'}\otimes y'_{b'}\mathcal{E}\otimes \mathcal{U}(\rho_{AB})]\nonumber\\
   &= \mathrm{tr}[\mathcal{E}^{\dagger}(x'_{a'})\otimes\mathcal{U}^{\dagger}( y'_{b'})\rho_{AB}]\nonumber,
\end{align}
where $\mathcal{E}^{\dagger}$ and $\mathcal{U}^{\dagger}$ are the dual channels of two given channels. Since $\rho_{AB}$ admit LHS model for any POVM measurements, \emph{viz.}, there exists $\{q(\lambda), \rho_{\lambda}\}$ such that
\begin{align}\label{}
p(a',b'|x',y') &=\int q(\lambda) p(a'|\mathcal{E}^{\dagger}(x'_{a'}),\lambda)\mathrm{tr}[\mathcal{U}^{\dagger}(y'_{b'})\rho_{\lambda}]d\lambda\nonumber\\
&=\int q(\lambda) p(a'|\mathcal{E}^{\dagger}(x'_{a'}),\lambda)\mathrm{tr}[ y'_{b'}\mathcal{U}(\rho_{\lambda})]d\lambda .\nonumber
\end{align}
Thus we can take LHS for $\tilde{\rho}_{A'B'}$ as $\{q(\lambda), \mathcal{U}(\rho_{\lambda})\}$, notice that quantum channel preserves the unitary operator, we have $\int \mathcal{U}(\rho_{\lambda}) d\lambda= \mathcal{U}(\mathds{1}_{\mathcal{K}})=\mathds{1}_{\mathcal{K}'}$, $\mathcal{U}(\rho_{\lambda})$ do be a set of local hidden states.\qed
\end{pf}
\end{lemma}
Note that the above result can be extend to the invertiable local operations and shared randomness (LOSR) transformations \cite{Beckman2001,Buscemi2012}, i.e. convex combination of some product channels $\Lambda=\sum_i p_i \mathcal{E}_i\otimes \mathcal{U}_i$. With this proposition we obtain the following result.

\begin{lemma}\label{prop:2}
 Given a set a bipartite states $\{\rho^{\eta}_{AB}\}_{\eta\in \mathbb{R}}\subset B(\mathcal{H}\otimes \mathcal{K})$ with the entangled, steerable, nonlocal ranges $R_E$, $R_S$, and $R_{NL}$ respectively, i.e., if $\eta \in R_E$, then $\rho^{\eta}_{AB}$ is a entangled state, similarly for the other two. Then under the invertible product channel $\mathcal{E}\otimes \mathcal{U}$ (i.e., there exist $\mathcal{E}^{-1}\otimes \mathcal{U}^{-1}$ such that $\mathcal{E}^{-1}\otimes \mathcal{U}^{-1}\circ \mathcal{E}\otimes \mathcal{U}$ is identity transformations), the states $\{\tilde{\rho}_{A'B'}^{\eta}=\mathcal{E}\otimes \mathcal{U}(\rho_{AB}^{\eta})\}_{\eta\in \mathbb{R}}$ have the same entangled (steerable, nonlocal) range as $\rho^{\eta}_{AB}$ .
\begin{pf} If $\rho_{AB}^{\eta}$ is separable, i.e., $\eta\not \in R_E$, then by prop \ref{prop:1}, $\tilde{\rho}_{A'B'}^{\eta}$ is separable, $\eta\not \in \tilde{R}_E$, thus we have $\tilde{R}_E\subseteq R_E$. Since the inverse channel  $\mathcal{E}^{-1}\otimes \mathcal{U}^{-1}$ is also a product channel, using $\rho_{AB}^{\eta}=\mathcal{E}^{-1}\otimes \mathcal{U}^{-1}(\tilde{\rho}_{A'B'}^{\eta})$ we get $R_E\subseteq\tilde{R}_E$, thus $R_E=\tilde{R}_E$. In the same way we can prove $R_S=\tilde{R}_S$ and $R_{NL}=\tilde{R}_{NL}$.
\qed
\end{pf}
\end{lemma}

We are now ready to construct states which are GME but not GMS. Consider Werner states acting on $\mathbb{C}^d\otimes \mathbb{C}^d$:
\begin{equation}\label{eq:werner}
W^{\eta}_{AB}=\eta \frac{2P_{as}}{d(d-1)}+(1-\eta)\frac{I_{d^2}}{d^2},
\end{equation}
where $P_{as}$ denotes the projector onto antisymmetric subspaces. The range for $W_{\eta}$ to be entangled but unsteerable under POVMs is $\frac{1}{(d+1)} < \eta <\frac{(3d-1)(d-1)^{d-1}(d^{-d})}{d+1}$ \cite{wener1989quantum,Barrett2002,Quintino2015}. Using the invertible channels $\mathcal{E}$ (resp. $\mathcal{U}$) which send $|i\rangle$ to $|i\rangle^{\otimes k}$ (resp. $|i\rangle^{\otimes l}$) for $i=1,\cdots, d$, we get the states
\begin{align}\label{eq:N-werner}
\tilde{W}^{\eta}_{A'B'}=&\eta \frac{\sum_{i<j} |\Psi_{ij}\rangle \langle \Psi_{ij}|}{d(d-1)} \nonumber \\
&+\frac{1-\eta}{d^2} (\sum_{i=1}^d (|i\rangle \langle i|)^{\otimes l})\otimes \sum_{i=1}^d (|i\rangle \langle i|)^{\otimes k}),
\end{align}
where $|\Psi_{ij}\rangle=(|i\rangle^{\otimes k}|j\rangle^{\otimes l}+|j\rangle^{\otimes k}|i\rangle^{\otimes l})/\sqrt{2}$,  $N=k+l$, $A'$ are the first $k$ particles and $B'$ are second $l$ particles. From proposition \ref{prop:1}, we know that the state $\tilde{W}^{\eta}_{A'B'}$ is not a GMS state if $\frac{1}{(d+1)} < \eta <\frac{(3d-1)(d-1)^{d-1}(d^{-d})}{d+1}$ , since it has LHS model with regard to the bipartition $A'|B'$. Note that the state is symmetric in its subsystem $A'$ and $B'$, and it is entangled with respect to the bipartition $A'|B'$, i.e., $P_{sym}^{A'}\otimes P_{sym}^{B'}\tilde{W}^{\eta}_{A'B'} P_{sym}^{A'}\otimes P_{sym}^{B'}= \tilde{W}^{\eta}_{A'B'}$ , using the criterion of GME state given in \cite{Augusiak2015}, which states that if $\rho_{A'B'}$ is entangled with respect to bipartition $A'|B'$ and symmetric for $A'$ and $B'$, then $\rho_{A'B'}$ is a GEM state, we get that $\tilde{W}^{\eta}_{A'B'}$ is a GME state, thus $\tilde{W}^{\eta}_{A'B'}\in \mathcal{S}_{GME}^{d,N}-\mathcal{S}_{GMS}^{d,N}$, GME and GMS are inequivalent. Although, we have constructed the example from Werner states, it is worth mentioning that the construction can also be applied to any entangled and unsteerable (under POVMs) states which are symmetric for parties in $A'$ and parties in $B'$.

One natural question is how to construct GMS states which are not contained in GMNL for any number of parties. But unfortunately, there is no good existing bipartite examples which is steerable but not Bell nonlocal.  There are some results for constructing LHV for entanglement states \cite{Augusiak2015,Augusiak2014,Hirsch2016,Cavalcanti2016} and construct LHS for entangled states \cite{Hirsch2016,Cavalcanti2016}, but very little is know for constructing steerable  local states. Since our method strongly depends on bipartite results, we can not using this method to construct GMS but not GMNL state, all we can  say is that $\mathcal{S}_{GMNL}^{d,N}\subseteq \mathcal{S}_{GMS}^{d,N}$. This is because that if a state admits a hybrid LHS model, then it must admits LHV model for arbitrary measurements, i.e., if $\rho$ is not GMS then it must not be GMNL. Above discussion can be summarized as the following proposition:

\begin{prop}
For any number $N$ and dimension $d$, we have $\mathcal{S}_{GMNL}^{d,N}\subseteq \mathcal{S}_{GMS}^{d,N}\subsetneq\mathcal{S}_{GME}^{d,N}$.
\end{prop}

\emph{Inequivalences of GMT, GMD and GME.}\textemdash
To define the genuine multipartite total (classical, discord) correlation, we first give some related definitions. A $k$-product state of $N$-partite system is of the form $\sigma=\sigma^{[1]}\otimes\cdots\otimes \sigma^{[m]}$ where each $\sigma^{[i]}$ is a $k_i$-partite state, $\sum_{i}k_i=N$ and $k_i \leq k$ for all $i=1,\cdots,m$. The set of all $k$-product state is denoted as $\EP_k$, it is not convex. The set of all $k$-classical state $\EC_k$, contains all states of the form $\chi=\sum_{\alpha_1,\cdots,\alpha_m}p_{\alpha_1,\cdots,\alpha_m}|\alpha_1\cdots \alpha_m\rangle\langle \alpha_1\cdots \alpha_m|=\sum_{\alpha}p_{\alpha}|\alpha\rangle \langle \alpha|$, where  local states $|\alpha_i\rangle\langle \alpha_i|$ span an orthogonal basis of $k_i$-partite subsystem, $\sum_{i}k_i=N$ and $k_i \leq k$ for all $i=1,\cdots,m$. $\EC_k$ is not convex, that is, mixing two classical states may give rise to a nonclassical state. The $k$-separable state set $\ES_k$ is the convex hull of $\EP_k$, viz, each state in $\ES_k$ is just some probabilistic mixture of states in $\EP_k$. Note that there are several important chain relations of these sets:
\begin{equation}\label{}
\begin{split}
&\EP_k \subseteq \EC_k \subseteq \ES_k, \,\,\,\,\,\, \forall k, \\
&\EP_1\subseteq \EP_2\subseteq \cdots\subseteq\EP_{N-1}\subseteq \EP_N, \\
&\EC_1\subseteq \EC_2\subseteq \cdots\subseteq\EC_{N-1}\subseteq \EC_N, \\
&\ES_1\subseteq \ES_2\subseteq \cdots\subseteq\ES_{N-1}\subseteq \ES_N.
\end{split}
\end{equation}

Taking similar approach as in \cite{Girolami2017}, we can quantify the genuine $k$-correlation by choosing a function $D$, which: (i) is non-negative, $D(\rho,\sigma)\geq 0$, and $D(\rho,\sigma)=0\Leftrightarrow \rho=\sigma$; (ii) is contractive under TPCP maps $\mathcal{E}$, viz, $D(\mathcal{E}(\rho),\mathcal{E}(\sigma))\leq D(\rho,\sigma)$. Genuine total correlation of order higher than $k$ can then been quantified as
\begin{equation}\label{}
D^{> k}_{tot}(\rho):=\min_{\sigma\in \EP_k}D(\rho,\sigma).
\end{equation}
If $D^{> k}_{tot}(\rho)>0$ we say that $\rho$ exhibiting genuine total correlation of order higher than $k$. The genuine quantum discord and genuine quantum entanglement can be characterized in the same way when we substitute $\EP_k$ with $\EC_k$ and $\ES_k$ respectively.

Here we only discuss the case for genuine multipartite ($N$-partite here) total (GMT) [resp. discord (GMD), entanglement (GME)] correlation, i.e., the minimum is taken over $\EP_{N-1}$, $\EC_{N-1}$ and $\ES_{N-1}$ respectively. It worth mentioning that $D_{ent}^{>N-1}(\rho)>0$ is equivalent to the failure of the hybrid separable model for state $\rho$.  To simplify the discussion, we choose relative entropy
\begin{equation}\label{}
\begin{split}
H(\rho||\sigma)=\left\{
\begin{array}{lcl}
\mathrm{tr}\rho \log \rho -\mathrm{tr}\rho \log \sigma, &\mathrm{supp}\rho\subseteq \mathrm{supp}\sigma\\
\infty, &\textrm{otherwise}
\end{array}\right.
\end{split}
\end{equation}
as the distance function. As has been proved in \cite{Modi2010,Szalay2015}, the closest state to $\EP_k$ is the product of its marginals, i.e., $\mathrm{argmin}\{H(\rho||\sigma):\sigma\in \EP_{k}\}=\rho^{[1]}\otimes\cdots\otimes \rho^{[m]}$ and $D^{>k}_{tot}(\rho)=\sum_i H(\rho^{[i]})-H(\rho)$, where each $\rho^{[i]}$ is $k_i$-particle reduced state of $\rho$.

\begin{prop}
For any number $N$ and dimension $d$, we have $\mathcal{S}_{GME}^{d,N}\subseteq \mathcal{S}_{GMD}^{d,N}\subseteq\mathcal{S}_{GMT}^{d,N}$.
\begin{pf}Since $\EP_{N-1} \subseteq \EC_{N-1} \subseteq \ES_{N-1}$, using the definition, we have $D^{> N-1}_{ent}(\rho) =\min_{\sigma\in \ES_{N-1}}D(\rho,\sigma)\leq D^{> N-1}_{dis}(\rho) =\min_{\sigma\in \ES_{N-1}}D(\rho,\sigma) \leq D^{> N-1}_{tot}(\rho) =\min_{\sigma\in \ES_{N-1}}D(\rho,\sigma)$. If a state $\rho \in \mathcal{S}_{GME}^{d,N}$, then $0<D^{>N-1}_{ent}\leq D^{>N-1}_{dis}$, therefore $\mathcal{S}_{GME}^{d,N}\subseteq \mathcal{S}_{GMD}^{d,N}$. The left part can be proved similarly. \qed
\end{pf}
\end{prop}

To prove the strict containing relation, we need to construct some states fall in between each two sets. But this is very difficult, now using proposition \ref{prop:1} we construct some GME but not GMT states. We start with the bipartite isotropic state $\rho^{\eta}_{AB}=\eta|GHZ \rangle \langle GHZ|+(1-\eta)I_{d^2}/d^2$ which is unentangled if $\eta \leq 1/(d+1)$ \cite{Horodecki1999}. Its steering range is $R_S=(\frac{H_d-1}{d-1},1]$ where $H_d=\sum_{n=1}^d1/n$ is harmonic series \cite{Wiseman2007}, nonlocal range is $R_{NL}=(\frac{(3d-1)(d-1)^{d-1}}{d^d(d+1)},1]$ \cite{Massar2002,Collins2002,Acin2006}. By choosing the same invertible channel as constructing GME but not GMS state in Eq. (\ref{eq:N-werner}) which maps $|i\rangle$ into $|i\rangle^{\otimes k}$ and $|i\rangle^{\otimes l}$, we obtain the state $\tilde{\rho}^{\eta}_{A'B'}=\eta |GHZ \rangle \langle GHZ|+(1-\eta) \tilde{I}_{A'}\otimes \tilde{I}_{B'}$, where $A'$, $B'$ contains $k$ and $l$ particles of the $N=k+l$ particle system, $\tilde{I}_{A'}=\sum_{i=1}^d|i\rangle \langle i|^{\otimes k}$ and $\tilde{I}_{B'}=\sum_{i=1}^d|i\rangle \langle i|^{\otimes l}$. From proposition \ref{prop:1}, if $\eta \leq 1/(d+1)$, $\tilde{\rho}^{\eta}_{A'B'}$ admits a separable model with respect to $A'|B'$ partition of $N$-particles. To make the analysis more easy, we mix some such kind of states together to obtain an $N$-partite state which is highly symmetric with respect to each subsystems
\begin{equation}\label{eq:N-isotropic}
\begin{split}
\tilde{\rho}^{\eta}&=\sum_{A'B'} \frac{1}{2^{N-1}-1}\tilde{\rho}^{\eta}_{A'B'}\\
&=\eta |GHZ\rangle \langle GHZ|+\sum_{A'B'} \frac{1-\eta}{2^{N-1}-1}\frac{\tilde{I}_{A'}}{d}\otimes \frac{ \tilde{I}_{B'}}{d},
\end{split}
\end{equation}
where summation is taken over all bipartitions $A'|B'$. The sate $\tilde{\rho}^{\eta}$ admits a hybrid separable model when $\eta\leq 1/(d+1)$ from proposition \ref{prop:1}, thus they are not GME, viz, $D^{>N-1}_{ent}(\tilde{\rho}^{\eta})=0$. We are now to show that such kind of states exhibit non-zero genuine total correlations. All detailed calculation is shown in supplemental material \cite{Supp}. Here we only give the result, there exist some $\eta \leq 1/(d+1)$ such that $D^{>N-1}_{tot}(\tilde{\rho}^{\eta})>0$. Thus we have $\mathcal{S}_{GME}^{d,N}\subsetneq\mathcal{S}_{GMT}^{d,N}$. For example, when $d=2$, $\eta=1/4$ and $N=3$, $D^{>3-1}_{tot}(\tilde{\rho}^{\frac{1}{4}})=0.337$, but $D^{>3-1}_{ent}(\tilde{\rho}^{\frac{1}{4}})=0$. In \cite{Giorgi2011}, tripartite case is discussed, the result is consistent with ours.

\emph{Conclusions.}\textemdash We have investigated the relation between GMT, GMD, GME, GMS and GMNL. By constructing states which are GMT but not GME, we show that GMT and GME are definitely different notions for any number of parties. Similarly, we show the inequivalence of GME and GMS. This gives an almost complete hierarchy of GM$\mathcal{C}$. However, there are still many challenging problems left open, e.g., is there any systematic approach to construct steerable local (in Bell scenario) states? is GMS and GMNL are the same notions or not? Similar questions also arise for investing the relation between GMD and GME. These interesting topics are all left for our future exploration.

\begin{acknowledgments}
We acknowledge Q.-Y. He and S.-L. Luo for many beneficial discussions.  This work is supported by and the National Natural Science Foundation of China (Grants No. 11275182 and No. 61435011), the National Key R \& D Program (Grant No. 2016YFA0301700), and the Strategic Priority Research Program of the Chinese Academy of Sciences (Grants No. XDB01030100 and No. XDB01030300).
\end{acknowledgments}

\appendix
\section{Supplemental material}
Here we give the detailed calculation of the state $\tilde{\rho}^{\eta}$ constructed from isotropic state, and we prove that if $\eta \leq 1/(d+1)$, i.e., $\tilde{\rho}^{\eta}\not \in \mathcal{S}^{d,N}_{GME}$, it can still have $D^{>N-1}_{tot}(\tilde{\rho}^{\eta})>0$, i.e., $\tilde{\rho}^{\eta} \in \mathcal{S}^{d,N}_{GMT}$. Therefore $\mathcal{S}^{d,N}_{GMT}\supsetneq \mathcal{S}^{d,N}_{GME}$.

Firstly, the state we constructed is
\begin{align}\label{}
\tilde{\rho}^{\eta}&=\sum_{A'B'} \frac{1}{2^{N-1}-1}\tilde{\rho}^{\eta}_{A'B'}\nonumber\\
&=\eta |GHZ\rangle \langle GHZ|+\sum_{A'B'} \frac{1-\eta}{2^{N-1}-1}\frac{\tilde{I}_{A'}}{d}\otimes \frac{ \tilde{I}_{B'}}{d}.
\end{align}
To write down the precise matrix form, we first label the basis $|0\rangle^{N}$, $|1\rangle^{N}$, $\cdots$, $|d-1\rangle^{N}$ as $v_1,\cdots,v_d$, label $|j\rangle^{k}\otimes |j\rangle^{l}$  with $k+l=N$ by $v_{d+1},\cdots,v_{d+(d^2-d)(2^{N-1}-1)}$, and all others are labeled by $v_{d+(d^2-d)(2^{N-1}-1)+1},\cdots,v_{d^N}$ in the increasing order of base-$d$ system of numbers. $\tilde{\rho}^{\eta}$ is then of the form
\begin{equation}\label{eq:state}
\tilde{\rho}^{\eta}=\left(
  \begin{array}{cccccccccccc}
  a & c & c &\cdots &c &0 & 0 &\cdots &0 &0&\cdots &0\\
  c & a & c &\cdots &c &0 & 0 &\cdots &0 &\vdots&\ddots& \vdots\\
  c & c & a &\cdots &c &0 & 0 &\cdots &0 &0&\cdots &0\\
  \vdots & \vdots & \vdots & \ddots &\vdots &\vdots &\vdots &\vdots &\vdots &\vdots&\ddots& \vdots\\
   c & c&c & \cdots &a & 0 & 0 &\cdots&0&0&\cdots &0\\
   0 & 0&0 & \cdots &0&b  &0 &\cdots &0&0&\cdots &0\\
   0 & 0&0 & \cdots &0&0  &b &\cdots &0&0&\cdots &0\\
   \vdots & \vdots&\vdots & \vdots &\vdots&\vdots &\vdots &\ddots &\vdots&\vdots&\ddots& \vdots\\
   0 & 0&0 & \cdots &0&0  &0&\cdots&b &0&\cdots &0\\
   0 & 0&0 & \cdots &0&0  &0 &\cdots &0&0&\cdots &0\\
  \vdots & \vdots&\vdots & \vdots &\vdots&\vdots &\vdots &\ddots &\vdots&\vdots&\ddots& \vdots\\
   0 & 0&0 & \cdots &0&0  &0 &\cdots &0&0&\cdots &0\\
  \end{array}
\right),
\end{equation}
where $a=\frac{\eta}{d}+\frac{1-\eta}{d^2}$, $b=\frac{1-\eta}{d^2 (2^{N-1}-1)}$ and $c=\frac{\eta}{d}$. We can check that $d a+(d^2-d)(2^N-1)b=1$. The eigenvalues of $\tilde{\rho}^{\eta}$ are then, $0$ of $d^N-d-(d^2-d)(2^{N-1}-1)$ degeneracy, $b$ of $(d^2-d)(2^{N-1}-1)$ degeneracy, $a-c$ of $(d-1)$ degeneracy , and $a+(d-1)c$ non-degenerate.

To calculate the genuine total correlation of $\tilde{\rho}^{\eta}$, we need to calculate
\begin{equation}\label{}
D^{>N-1}_{tot}(\tilde{\rho}^{\eta})=\min_{\tilde{\rho}^{[1]}_{\eta}\otimes \cdots \otimes\tilde{\rho}^{[m]}_{\eta}\in\EP_{N-1}}\{\sum_{i}H(\tilde{\rho}^{[i]}_{\eta})-H(\tilde{\rho}^{\eta})\}
\end{equation}
But note that (i) all $k$-partite reduced state of $\tilde{\rho}$ are of the same form, since $\tilde{\rho}$ is symmetric with respect to each particle; (ii) the sum of von Neumann entropies the reduced states is always greater than or equal to the overall state. Then $D^{>N-1}_{tot}(\tilde{\rho})= H(\tilde{\rho}^{A'}_{\eta})+H(\tilde{\rho}^{B'}_{\eta})-H(\tilde{\rho}^{\eta})$, $A'$ is an one-particle subsystem, $B'$ is the system of the rest $(N-1)$ particles. For convenience we let $A'$ be the first particle, by tracing out all other $N-1$ particles, the density matrix is $\rho_{\eta}^{A'}= \tilde{I}_{d^2}/d$. Similarly, the reduced matrix of $N-1$ particle subsystem $B'$ is
\begin{equation}\label{}
\rho_{\eta}^{B'}=\mathrm{diag}(a',\cdots,a',b',\cdots,b',0,\cdots,0),
\end{equation}
where $a'=\frac{\eta}{d}+\frac{1-\eta}{(2^{N-1}-1)d}+\frac{(2^{N-1}-2)(1-\eta)}{(2^{N-1}-1)d^2}$ and $b'=2b$, $b$ is the same as in Eq. (\ref{eq:state}). The number of $a'$ and $b'$ are $d$ and $(2^{N-2}-1)(d^2-d)$. We can also check that $d a'+(2^{N-2}-1)(d^2-d)b'=1$.

We are now in a position to give the precise formula of $D^{>N-1}_{tot}(\tilde{\rho})$, it is of the form:
\begin{align}\label{}
D^{>N-1}_{tot}(\tilde{\rho}^{\eta})=&H(\tilde{\rho}_{\eta}^{A'})+H(\tilde{\rho}_{\eta}^{B'})- H(\tilde{\rho}^{\eta})\nonumber  \\
=& \log d -d a'\log a'-(2^{N-2}-1)(d^2-d) b'\log b'\nonumber \\
&+ (d-1)(a-c)\log (a-c)\nonumber \\
&+ [a+(d-1)c]\log [a+(d-1)c]\nonumber \\
&+ (d^2-d)(2^{N-1}-1) b\log b.
\end{align}
When taking $\eta\leq 1/(d+1)$, $D^{>N-1}_{tot}(\tilde{\rho}^{\eta})$ can still be greater than zero, this complete the proof.
\bibliographystyle{apsrev4-1-title}

\end{document}